\documentclass[a4paper,11pt]{article}
\usepackage{jinstpub} 
\usepackage{lineno}
\usepackage{subcaption}

\usepackage{multirow} 
\usepackage{booktabs}    
\usepackage{array}
\usepackage{xcolor}

\usepackage{lmodern}

\title{\boldmath Simulation of radiation environment for the beam monitor of CEE experiment}

\author[a,b]{Qian Wang,}
\author[a,b,1]{Hulin Wang,\note{Corresponding author.}}
\author[a,b]{Chaosong Gao,}
\author[a,b]{Jun Liu,}
\author[c]{Xianglun Wei,}
\author[a,b]{Junshuai Liu,}
\author[d]{Zhen Wang,}
\author[a,b]{Ran Chen,}
\author[c]{Peng Ma,}
\author[c]{Haibo Yang,}
\author[c]{Chengxin Zhao,}
\author[a,b]{Mingmei Xu,}
\author[a,b]{Shusu Shi,}
\author[a,b]{Xiangming Sun,}
\author[a,b]{Feng Liu}

\affiliation[a]{PLAC, Key Laboratory of Quark \& Lepton Physics (MOE), Central China Normal University, Wuhan, 430079, China}
\affiliation[b]{Hubei Provincial Engineering Research Center of Silicon Pixel Chip \& Detection Technology, Wuhan, 430079, China}
\affiliation[c]{Institute of Modern Physics, Chinese Academy of Sciences, Lanzhou, 730000, China}
\affiliation[d]{School of Physics and Electronic Science, Guizhou Normal University, Guiyang, 550001, China}

\emailAdd{hulin.wang@ccnu.edu.cn}

\abstract{
The cooling storage ring external-target experiment is a large-scale nuclear physics experiment, which aims to study the physics of heavy-ion collisions at low temperatures and high baryon densities. 
A beam monitor (BM) is placed in the beam line to monitor the beam status and to improve the reconstruction resolution of the primary vertices.
The radiation dose and particle fluence stemming from the beam interactions with gases and detector materials affect the performance of the sensors and electronics of BM.
This paper uses FLUKA Monte Carlo code to simulate the radiation environment of BM detector. 
Radiation quantities including the total ionizing dose, 1 MeV neutron equivalent fluence, high-energy hadron flux, thermal neutron flux, and nuclear fragment flux are presented. 
Results of alternative simulation setups, including adding shielding layers inside the BM, are also investigated.}


\begin{document}
\maketitle
\flushbottom

\section{Introduction}
\label{sec:intro}
\begin{figure}[htbp] 
        \centering
         \begin{subfigure}{0.42\textwidth}
            \centering
            \includegraphics[width=\textwidth]{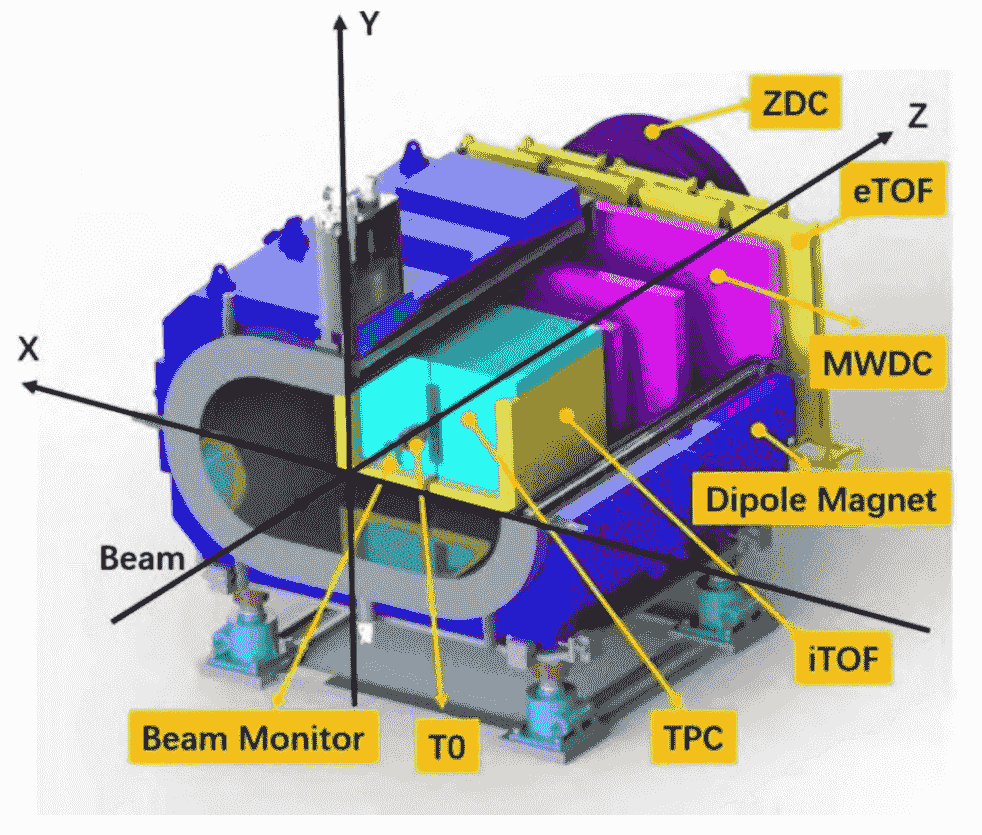}
            \caption{\label{subfig:struct}}
        \end{subfigure} 
        \quad
        \begin{subfigure}{0.54\textwidth}
            \centering
            \includegraphics[width=\textwidth]{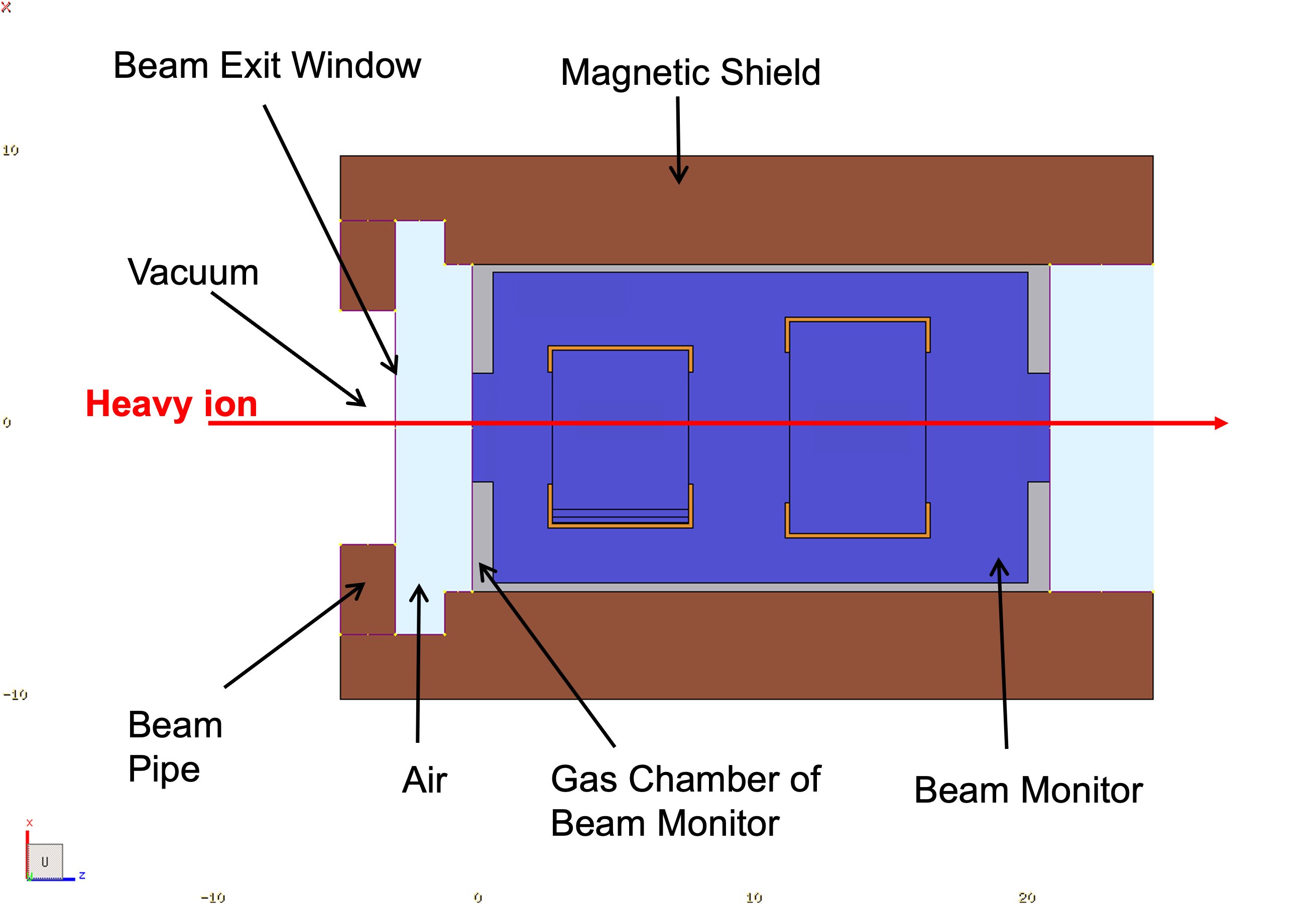}  
            \caption{\label{subfig:shieldBM}}
        \end{subfigure}
    \caption{~\label{fig:cee}The overall structure of the CEE spectrometer~\cite{Wang2022} (a), and the BM inside the magnetic shield in FLUKA (b).}
\end{figure}

The cooling storage ring (CSR)~\cite{YUAN2013214} external-target experiment (CEE)~\cite{Lü2016} at the heavy-ion research facility in Lanzhou (HIRFL) is a large-scale nuclear physics experiment, which aims to study the nuclear matter at low temperatures and high baryon densities~\cite{particles3020022}.
It is a fixed-target experiment utilizing heavy-ion beams provided by the HIFRL-CSR.
The overall structure of the CEE spectrometer is shown in Figure~\ref{subfig:struct}. 
It includes a dipole magnet, a time projection chamber (TPC)~\cite{YUAN2023168281}, multi-wire drift chambers (MWDCs)~\cite{Lyu2020}, 
a start time detector (T0)~\cite{Hu_2019}, an inner time-of-flight detector (iTOF)~\cite{Wang_2022}, an external time-of-flight detector (eTOF)~\cite{Wang_2023}, 
a zero-degree calorimeter (ZDC)~\cite{Zhu_2021}, and a beam monitor (BM)~\cite{Wang2022,Wang_2025}. 

As shown in Figure~\ref{subfig:shieldBM}, the BM is placed inside the magnetic shield in the beam line, downstream of the beam exit window and before the fixed target.
It is designed to measure the time and position of each beam particle at a maximum beam particle rate of 1 MHz, 
so as to monitor the beam status and to improve the reconstruction resolution of the primary vertices of the collisions.  
The design specifications mainly include a spatial resolution of better than 50 $\mu$m, and a time resolution of better than 1 $\mu$s. 
The BM is operated under the atmospheric pressure and the distance between each of the planes of sensor chips and the beam center is about 3.6 cm.
As a result, its sensors and electronics are expected to be impacted by the radiation environment, 
including the possible damages from total ionizing dose (TID) effects, single event effects (SEEs) and non-ionizing energy loss (NIEL) effects.

This paper uses the FLUKA~\cite{10.3389/fphy.2021.788253,BATTISTONI201510,refId0, refId01} Monte Carlo code to analyze the radiation environment of BM,
in particular in the planes of sensor chips. 
Section~\ref{sec:setup} describes the detector geometry, running scenario and MC modelling setups.
In Section~\ref{sec:result} the results of key radiation quantities are evaluated, 
including TID, NIEL expressed in 1 MeV neutron equivalent fluence, high-energy hadron flux, thermal neutron flux, and nuclear fragment flux.
In addition, the effect of adding radiation shielding layers inside the BM is studied, as well as other alternative simulation setups.
Finally, Section~\ref{sec:conc} presents the conclusions.

\section{Simulation setups}
\label{sec:setup}

\subsection{Geometry}
\label{subsec:geo}

\begin{figure}[htbp] 
        \centering
         \begin{subfigure}{0.56\textwidth}
            \centering        
            \includegraphics[width=\textwidth]{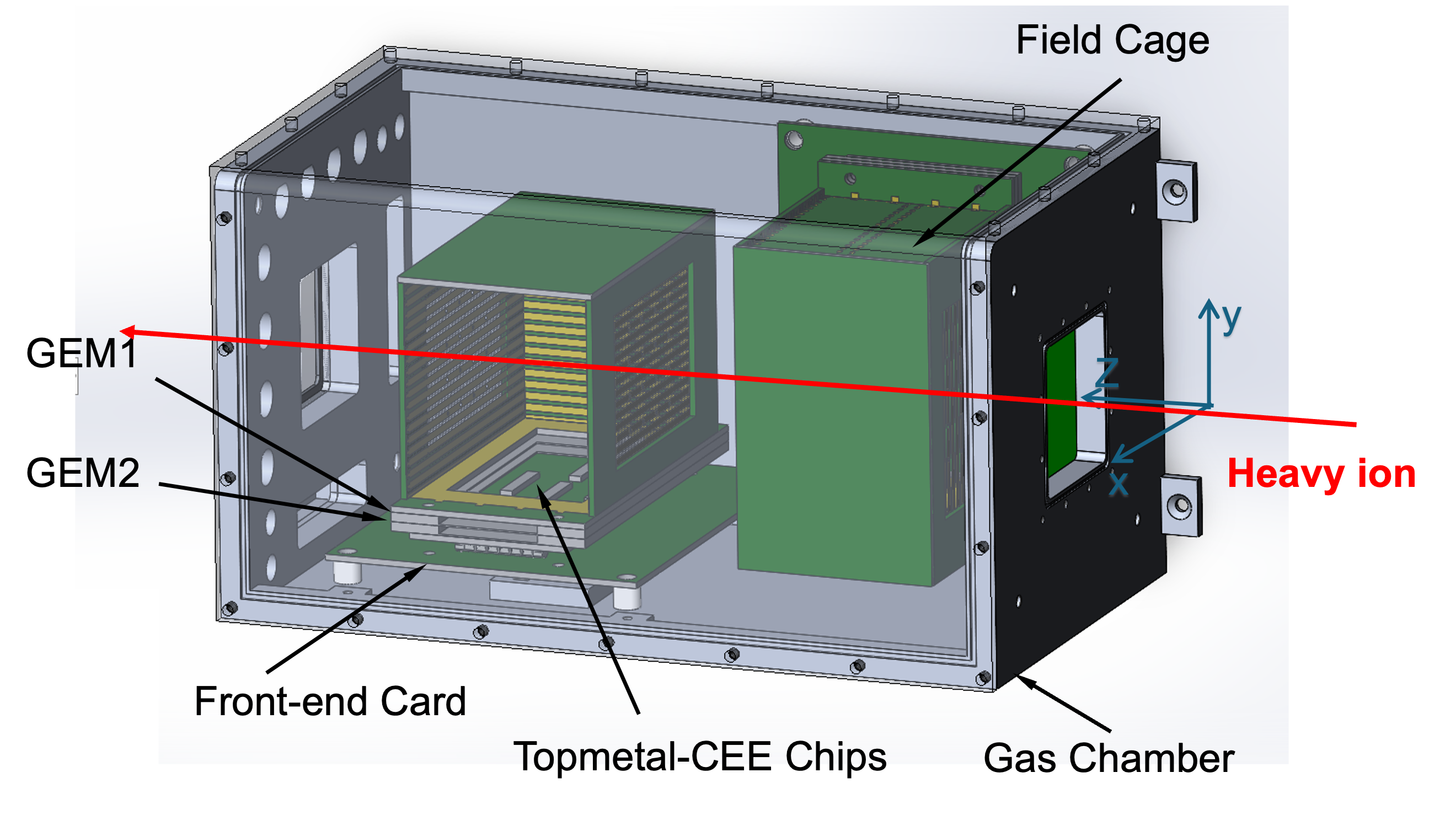}
            \caption{\label{fig:BMstruct}}   
        \end{subfigure} 
        \quad
        \begin{subfigure}{0.35\textwidth}
            \centering            
            \includegraphics[width=\textwidth]{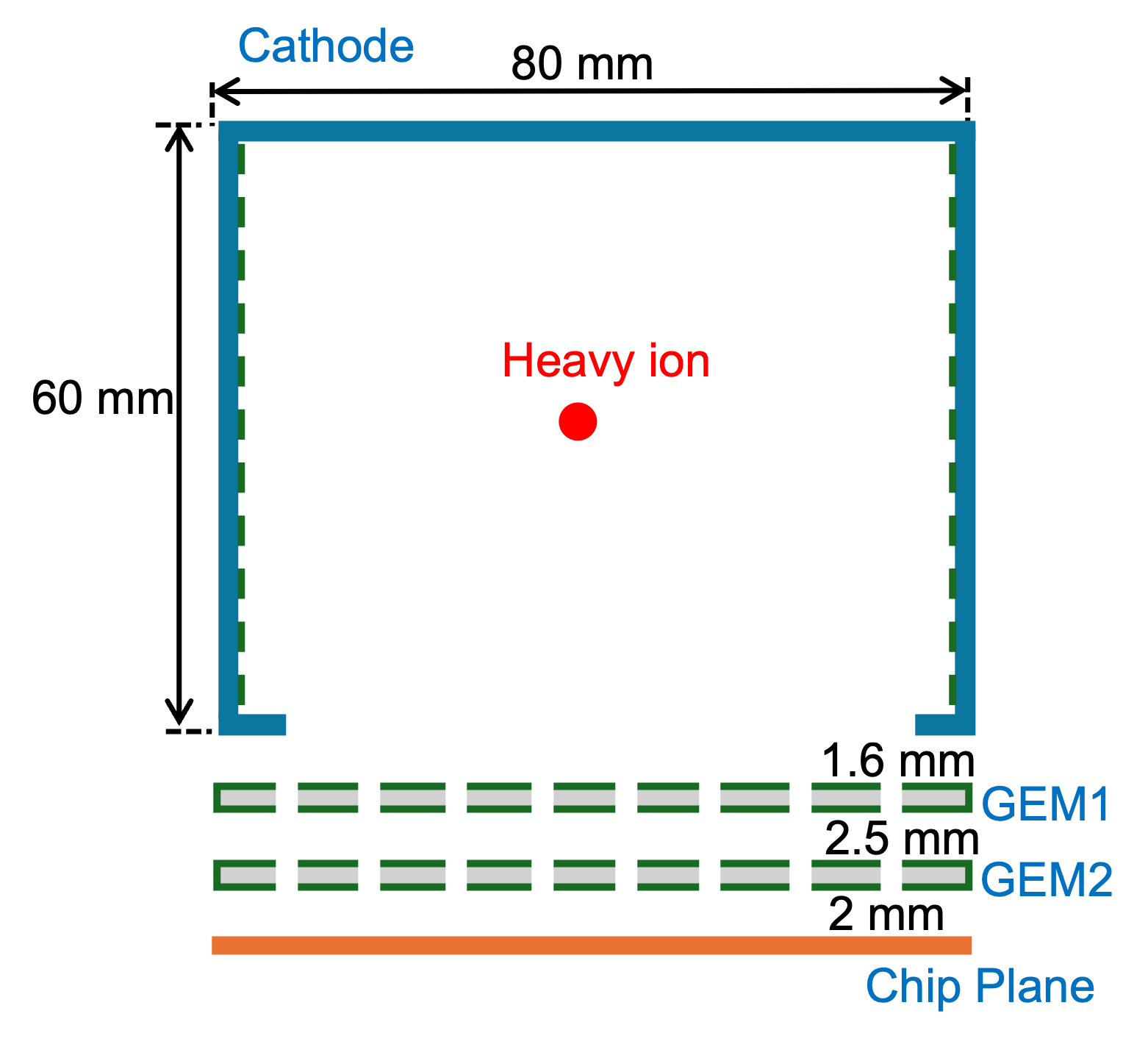} 
            \caption{\label{fig:fieldcage}}
        \end{subfigure}
    \caption{The structure of the beam monitor (a) and the schematic of its readout (b).}
\end{figure}

The structure of BM is shown in Figure~\ref{fig:BMstruct}.
It is placed 2.8 cm downstream of the beam exit window made of 28 $\mu$m thick titanium, with air in between.
In the simulation, detailed geometry and material descriptions are provided for the major components including 
gas chamber, two field cages, gas electron multipliers (GEMs), planes of Topmetal-CEE~\cite{LIU2023167786} sensor chips,
the beam exit window, and the gases within and outside of the BM at atmospheric pressure.
The front-end cards~\cite{Liu2025}, support structures, and cables all lie further away from the beam line than the chip planes, and are not included in the simulation. 

The gas chamber is made of aluminum, with dimensions of 120 ($x$) $\times$ 120 ($y$) $\times$ 212 ($z$) mm$^3$.
Throughout the paper, the right-handed coordinate system is used, with the $z$ axis coinciding with the nominal beam path and the $y$ axis pointing upward.
The thicknesses of the front/back and side walls are 8 mm and 3 mm, respectively. 
The chamber has an entrance/exit window of 40 ($x$) $\times$ 40 ($y$) mm$^2$, formed by a 2 $\mu$m thick Mylar film with 200 nm thick aluminized layers on both sides.
The chamber is filled with a gas mixture of Ar($70\%$) + CO$_2$($30\%$).

The two field cages have the same size of 60 (electric field direction) $\times$ 80 $\times$ 50 (beam direction) mm$^3$,
and are placed so that the electric fields are orthogonal to each other.
They are formed by 1.6 mm thick PCB boards, with cathodes and drift electrodes made of 35 $\mu$m thick copper cladded with 0.05 $\mu$m thick gold.
The planes of entrance and exit windows are perpendicular to the beam direction, both with dimensions of 41 (electric field direction) $\times$ 55 mm$^2$.
They are made of 25.4 $\mu$m thick Kapton films with drift electrodes made of 35 $\mu$m thick copper cladded with 0.05 $\mu$m thick gold.

Figure~\ref{fig:fieldcage} illustrates the schematic of the readout of BM. 
Zero to two GEM layers can be inserted between the field cage and the chip plane, 
to be able to maximize the detector performance for different ion species from C to U with ionizing powers spanning two orders of magnitude.
Two GEM layers are used as the benchmark,
and the GEM is modelled by 50 $\mu$m thick Kapton with 5 $\mu$m thick copper on both sides. 
The chip plane is modelled by 300 $\mu$m thick silicon, below which lies the 1.6 mm thick PCB board.

\subsection{Running scenario and MC modelling setups}
\label{subsec:runmc}

The CEE experiment plans to run for approximately two months per year for a minimum of three years.
The 500 MeV/u U-ions and 1 GeV/u C-ions are two typical heavy-ion beams and are used as benchmarks in the simulation. 
The beam profile is assumed to follow the 2-D Gaussian distribution, with a full width at half maximum (FWHM) of 2.35 mm used as the benchmark.
The beam particle rate is set to 1 MHz.

The set of default options called PRECISION is used in the simulation. 
For the physics of heavy-ion beams, three additional PHYSICS cards are used to enable 
the ion electromagnetic dissociation, coalescence mechanisms, and new FLUKA evaporation model with heavy fragment evaporation.
The transport and production thresholds for electrons and positrons are set to 10 keV, 
which corresponds to a range of 0.21 cm in the working gas.
These thresholds are set to 1 keV for photons, and set to the thermal energy of 296 K for neutrons.
The delta-ray production threshold is set to 10 keV.
For all the studies, $3\times10^{5}$ ($9\times10^{7}$) events are produced for U-ion (C-ion) beam, 
except for the studies in Section~\ref{subsubsec:cagev1}, where $9\times10^{5}$ events are produced for U-ion beam for the BM with less material.

\section{Simulation results}
\label{sec:result}

The TID and NIEL are related to the long-term performance degradation or damages to the chips, 
while the high-energy hadron flux, thermal neutron flux, and nuclear fragment flux determine the rate of SEEs.

The TID is defined as the amount of ionizing energy deposited per unit mass of material. 
It causes charge accumulations in the oxide layers and oxide-bulk interfaces, 
leading to the threshold voltage drifts and increased leakage currents to the electronic devices~\cite{1208572,CASSANI2021109338}.
The NIEL causes the displacement damages in the semiconductor materials, which can give rise to the changes of doping concentrations and increased leakage currents.
It is characterized by an energy-dependent damage function~\cite{4336006,8331152}, and is normally measured as 1 MeV equivalent neutron fluence.

High-energy hadrons induce SEEs through inelastic nuclear interactions. 
A lower energy threshold of 20 MeV is typically adopted, as the SEE cross section is approximately constant for hadrons above 20 MeV and decreases rapidly below it~\cite{10066303}.
Thermal neutrons are captured by the traces of $^{10}$B in the electronic devices, 
in each interaction producing an $\alpha$ particle and a Li ion, with sufficient linear energy transfers (LETs) to induce SEEs~\cite{HUHTINEN2000155,9102299}.
The nuclear fragments are produced by the interactions of primary ions or secondary particles with detector materials or gases.
They can induce the SEEs either through direct ionization in case of sufficient LETs or indirectly through nuclear interactions~\cite{10518067,8995568}.

\subsection{TID}
\label{subsec:tid}

\begin{figure}[htbp] 
        \centering
         \begin{subfigure}{0.48\textwidth}
            \centering
            \includegraphics[width=\textwidth]{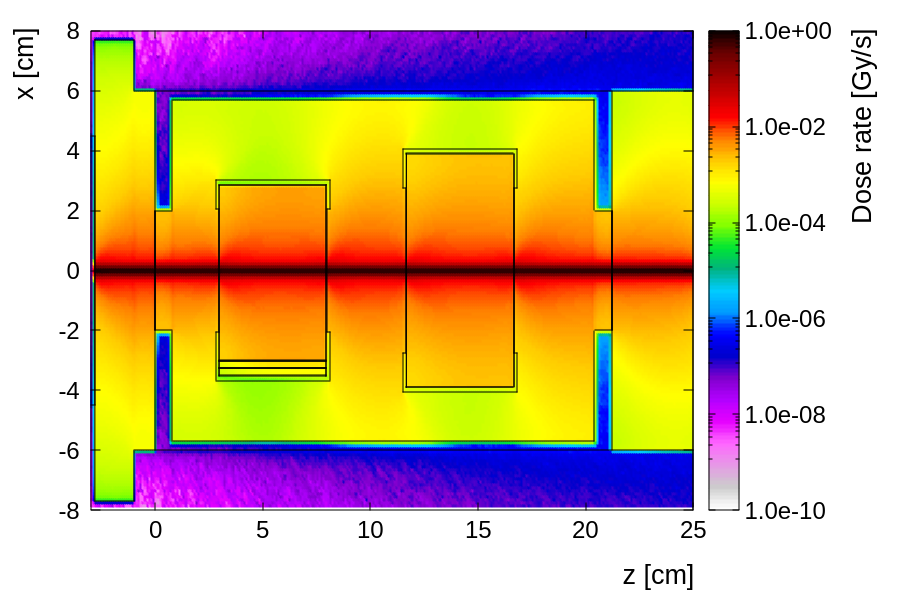}
            \caption{\label{subfig:UTID}}
        \end{subfigure} 
        \quad
        \begin{subfigure}{0.48\textwidth}
            \centering
            \includegraphics[width=\textwidth]{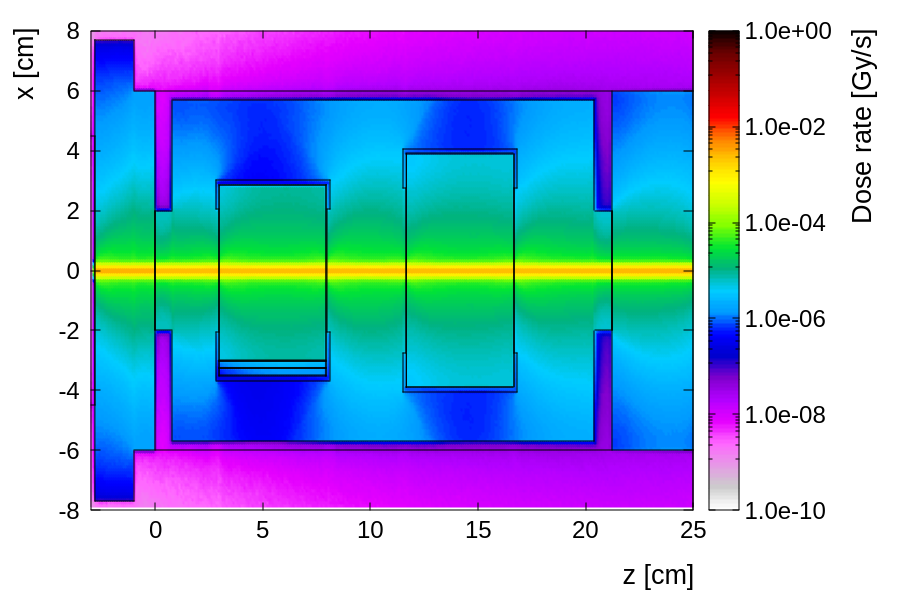}
            \caption{\label{subfig:CTID}}
        \end{subfigure}
    \caption{~\label{fig:mapdose} 
    Maps of the ionizing dose rates for U-ion (a) and C-ion (b) beams.
    }
\end{figure}

\begin{figure}[htbp] 
        \centering
        \begin{subfigure}{0.48\textwidth}
            \centering
            \includegraphics[width=\textwidth]{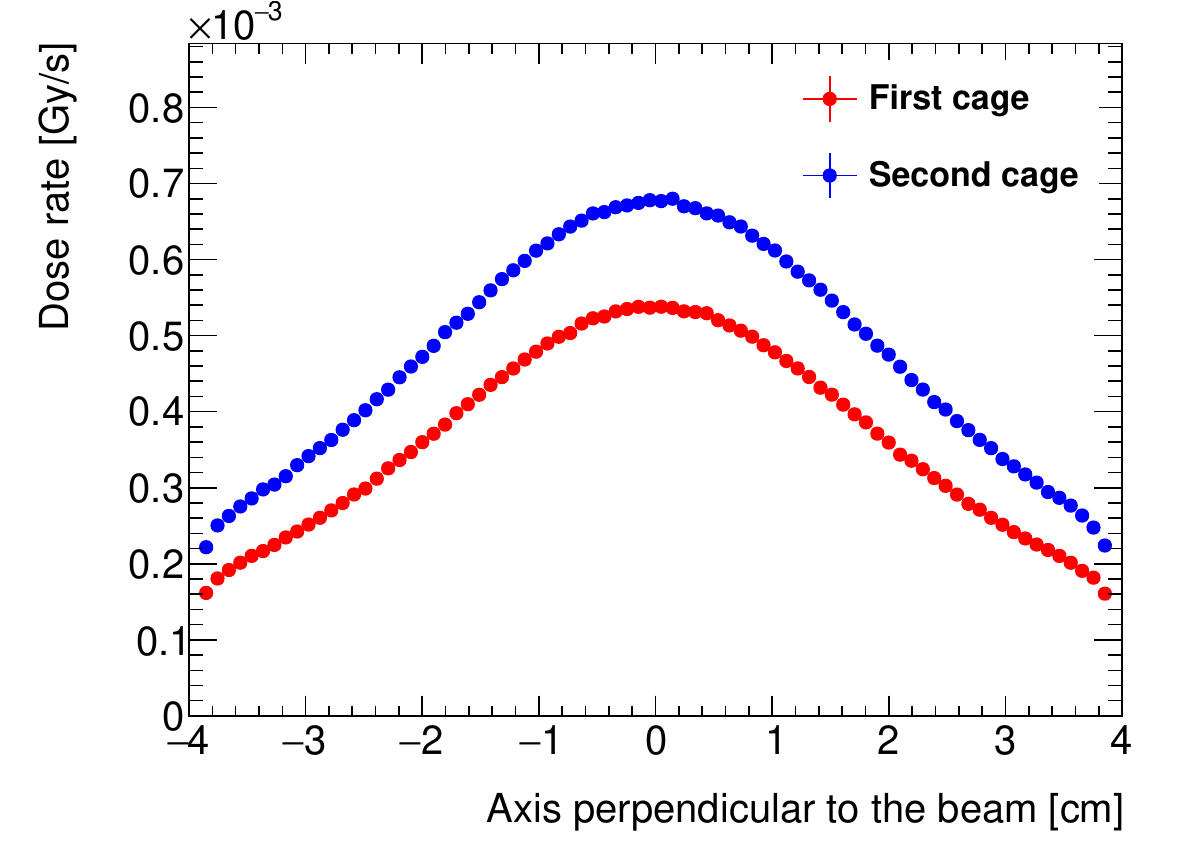}
            \caption{\label{subfig:UGEM2chip}}
        \end{subfigure}
        \quad
        \begin{subfigure}{0.48\textwidth}
            \centering
            \includegraphics[width=\textwidth]{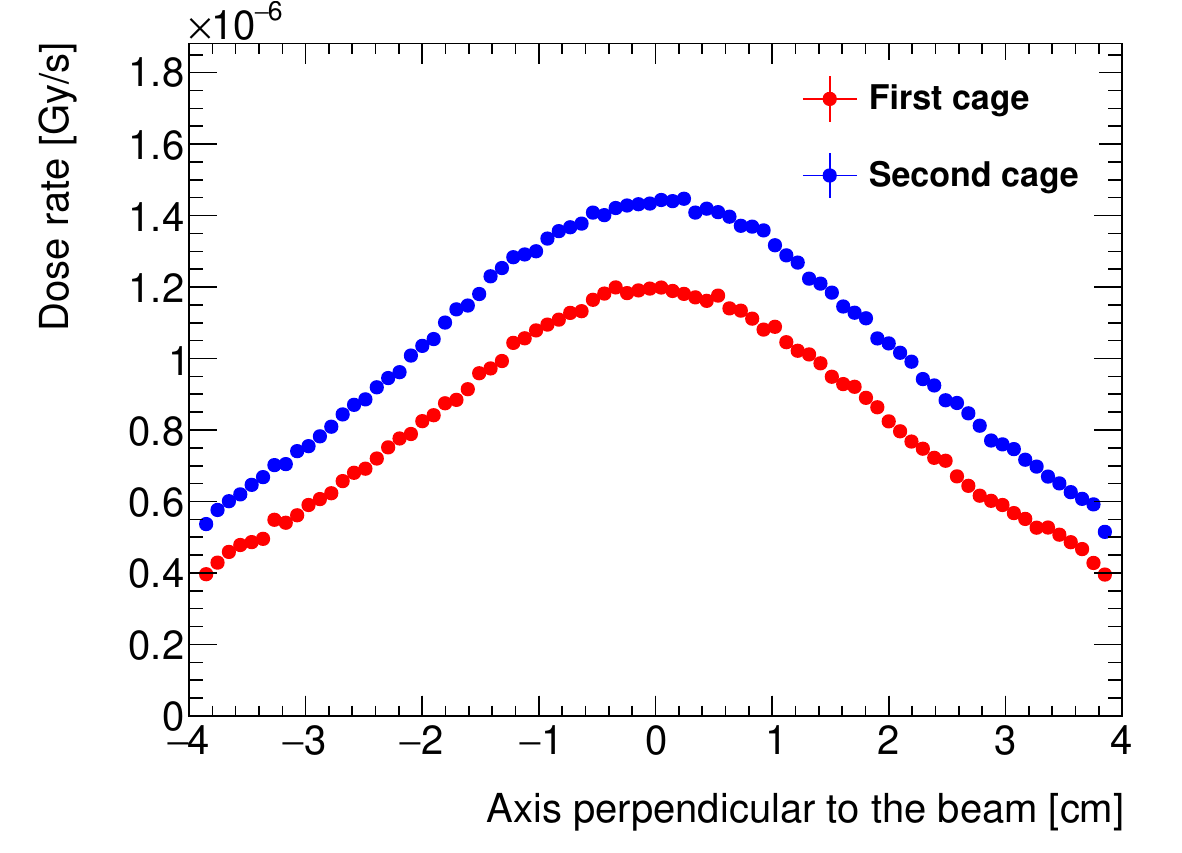}  
            \caption{\label{subfig:CGEM2chip}}
        \end{subfigure}
    \caption{~\label{fig:chipdose} 
    Dose rates in chip planes as a function of the coordinate of axis perpendicular to the beam, for U-ion (a) and C-ion (b) beams.}
\end{figure}

The maps of ionizing dose rates for U-ion and C-ion beams are shown in Figure~\ref{fig:mapdose}, 
in which the dose rates are averaged along the $y$ axis.
Figure~\ref{fig:chipdose} shows the dose rate distributions for the chip planes,
in which the dose rates are averaged along the $z$ axis (i.e. beam direction), and presented as a function of the coordinate of the axis perpendicular to the beam. 
The peak dose rates for the chip planes in the first cage (the upper stream one) and the second field cage are approximately 
$540$ ($1.2$) $\mu$Gy/s and $680$ ($1.4$) $\mu$Gy/s for U-ion (C-ion) beams, respectively.
The ionizing doses for the chip planes are predominantly contributed by electrons.
The other particles (such as neutrons, photons, and protons) in total contribute less than $0.5\%$ of the doses.

With the total expected beam time of 6 months in three years' running time, the TIDs in the chip planes are given in Table~\ref{tab:tid}.
The maximum value amounts to 10.6 kGy for the chip plane in the second field cage for the U-ion beam.  

\begin{table}[hbtp]
    \centering
    \caption{\label{tab:tid} The TIDs in the chip planes for a beam time of 6 months.
    The numbers correspond to the average values while the numbers in brackets refer to the peak values.}
    \begin{tabular}{lcc}
    \hline\hline
     & First cage [kGy] & Second cage [kGy] \\
    \hline
    U-ion   & $5.699 \pm 0.002$ ($8.367 \pm 0.023$) & $7.409 \pm 0.002$ ($10.575 \pm 0.025$)  \\
    C-ion   & $1.297 \pm 0.001$ ($1.865 \pm 0.016$) $\times 10^{-2}$ & $1.618 \pm 0.002$ ($2.251 \pm 0.016$) $\times 10^{-2}$  \\
    \hline\hline
    \end{tabular}
\end{table}

\subsection{NIEL}
\label{subsec:niel}

\begin{figure}[htbp]
        \centering
         \begin{subfigure}{0.48\textwidth}
            \centering
            \includegraphics[width=\textwidth]{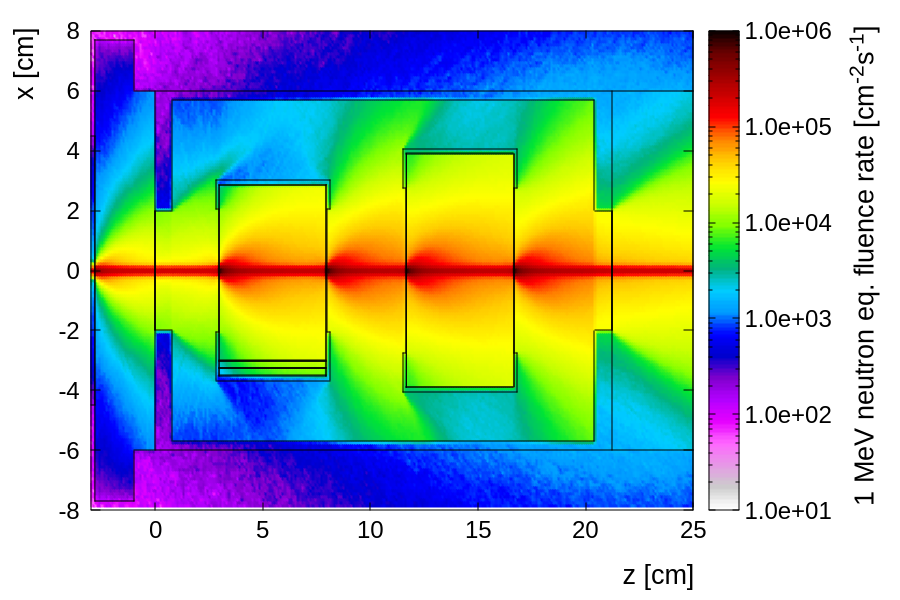}
            \caption{~\label{subfig:U1MeVSiBM}}
        \end{subfigure}
        \quad
        \begin{subfigure}{0.48\textwidth}
            \centering
            \includegraphics[width=\textwidth]{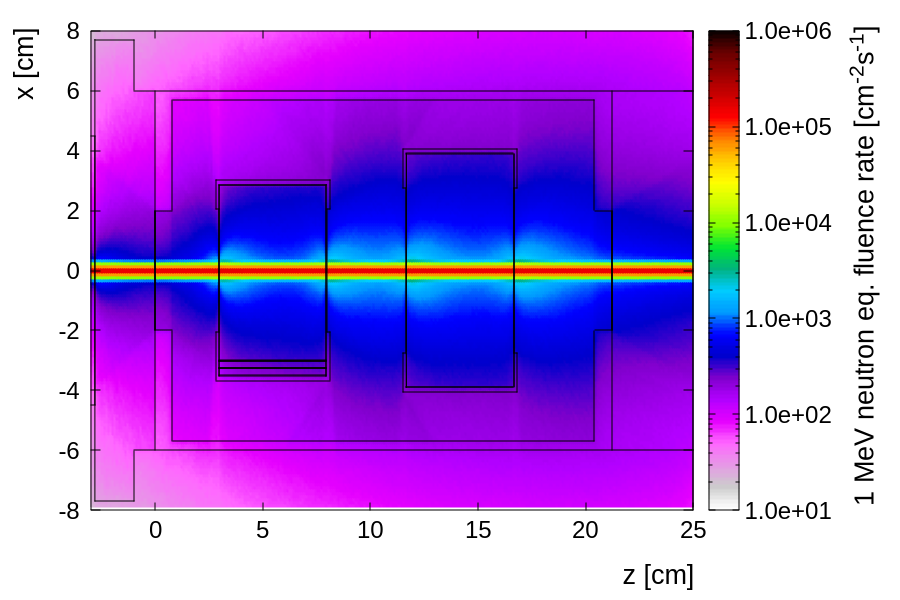}
            \caption{~\label{subfig:Usi1Mevchip}}
        \end{subfigure}
    \caption{~\label{fig:mapniel} Maps of the 1 MeV neutron equivalent fluence rates for U-ion (a) and C-ion (b) beams.}
\end{figure}

\begin{figure}[htbp]
        \centering
         \begin{subfigure}{0.48\textwidth}
            \centering
            \includegraphics[width=\textwidth]{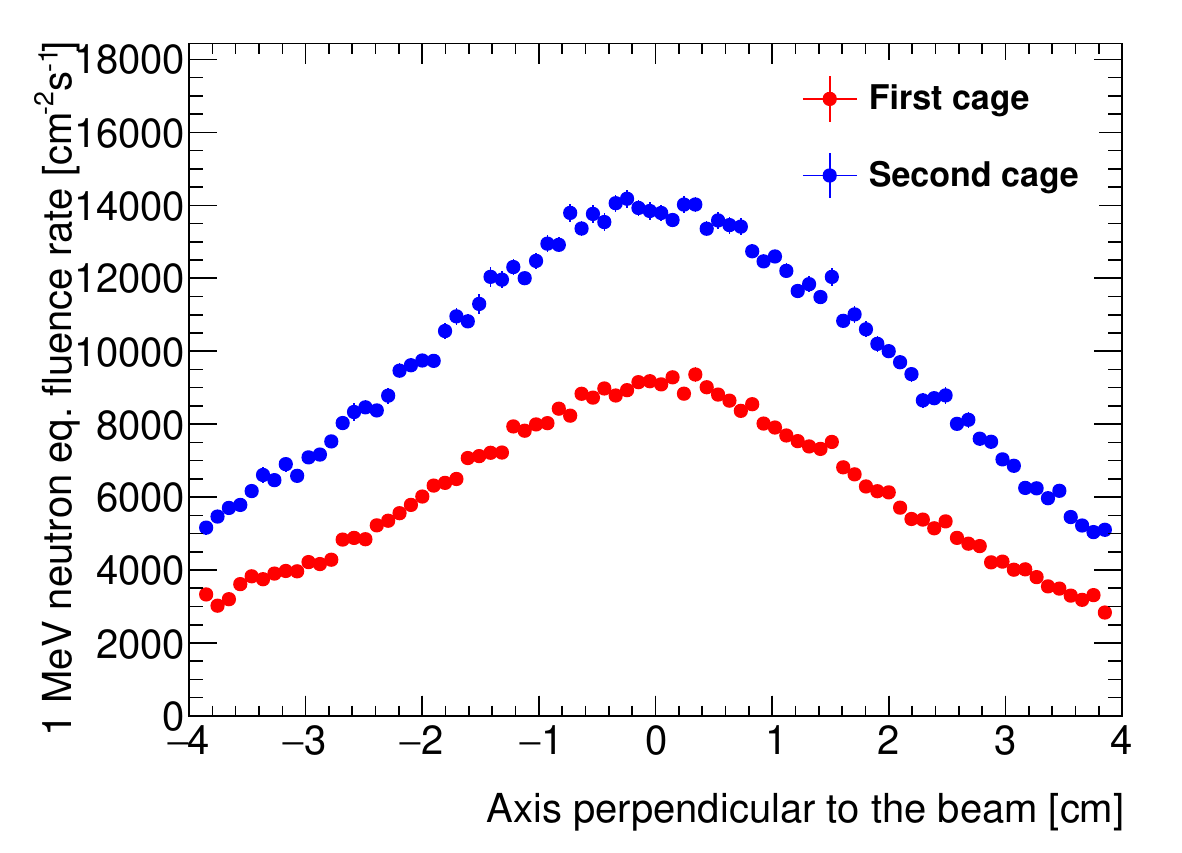}
            \caption{~\label{subfig:C1MeVSiBM}}
        \end{subfigure}
        \quad
        \begin{subfigure}{0.48\textwidth}
            \centering
            \includegraphics[width=\textwidth]{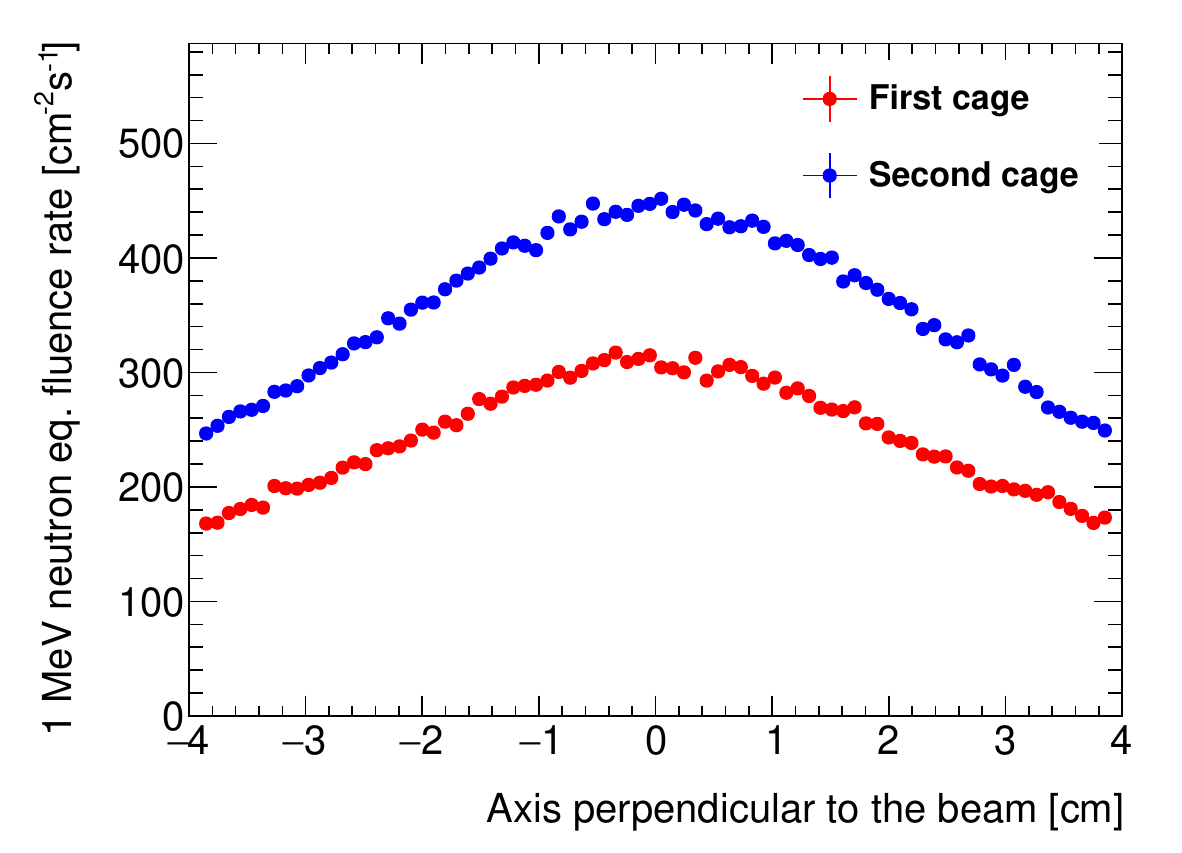}
            \caption{~\label{subfig:Csi1Mevchip}}
        \end{subfigure}
    \caption{~\label{fig:chipniel}  The 1 MeV neutron equivalent fluxes in chip planes as a function of the coordinate of axis perpendicular to the beam, for U-ion (a) and C-ion (b) beams.}
\end{figure}

Figure~\ref{fig:mapniel} shows the maps of NIELs in 1 MeV neutron equivalent fluence rates for U-ion and C-ion beams.
The NIELs in the chip planes are shown in Figure~\ref{fig:chipniel}.
The peak fluence rates for the chip planes in the first and second field cages are approximately 9400 (320) cm$^{-2}$s$^{-1}$ and 14000 (450) cm$^{-2}$s$^{-1}$
for U-ion (C-ion) beams, respectively.
The leading contributions in the chip planes are from electrons, which are approximately $80\%$.
The subleading contributions are from protons and neutrons, each constituting about $5\%$ to $10\%$.

The fluences are given in Table~\ref{tab:niel}. 
The maximum value is 2.2 $\times 10^{11}$ cm$^{-2}$, corresponding to the chip plane in the second field cage for the U-ion beam. 

\begin{table}[hbtp]
    \centering
    \caption{\label{tab:niel} The 1 MeV neutron equivalent fluence in the chip planes for a beam time of 6 months.
    The numbers correspond to the average values while the numbers in brackets refer to the peak values.}
    \begin{tabular}{lcc}
    \hline\hline
     & First cage [cm$^{-2}$] & Second cage [cm$^{-2}$] \\
    \hline
    U-ion   & $0.959 \pm 0.002$ ($1.456 \pm 0.033$) $\times 10^{11}$ & $1.541 \pm 0.004$ ($2.205 \pm 0.038$) $\times 10^{11}$  \\
    C-ion   & $3.838 \pm 0.007$ ($4.935 \pm 0.073$) $\times 10^{9}$ & $5.586 \pm 0.009$ ($7.025 \pm 0.084$) $\times 10^{9}$  \\
    \hline\hline
    \end{tabular}
\end{table}

\subsection{High-energy hadron flux}
\label{subsec:heh}

\begin{figure}[htbp] 
        \centering
         \begin{subfigure}{0.48\textwidth}
            \centering
            \includegraphics[width=\textwidth]{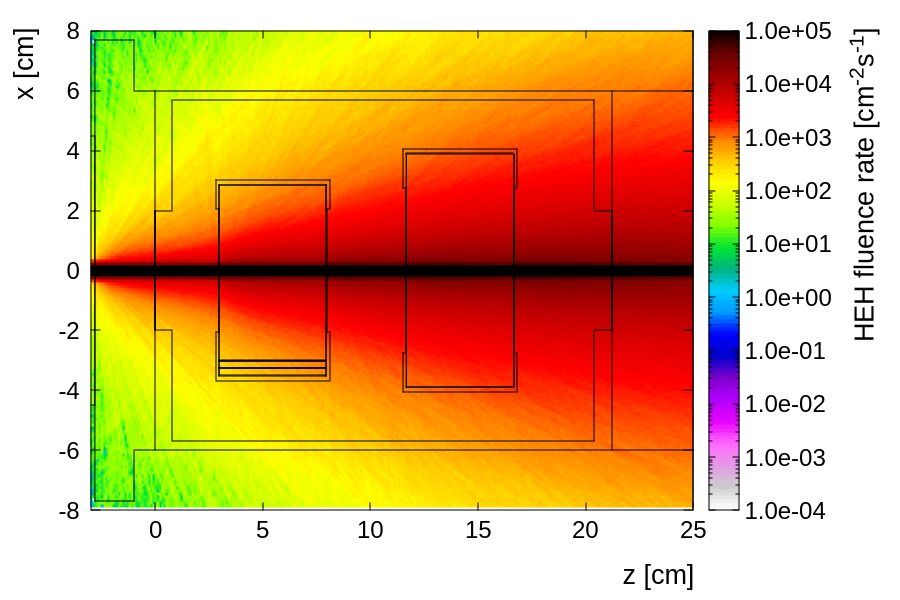}
            \caption{\label{subfig:maphehU}}
        \end{subfigure} 
        \quad
        \begin{subfigure}{0.48\textwidth}
            \centering
            \includegraphics[width=\textwidth]{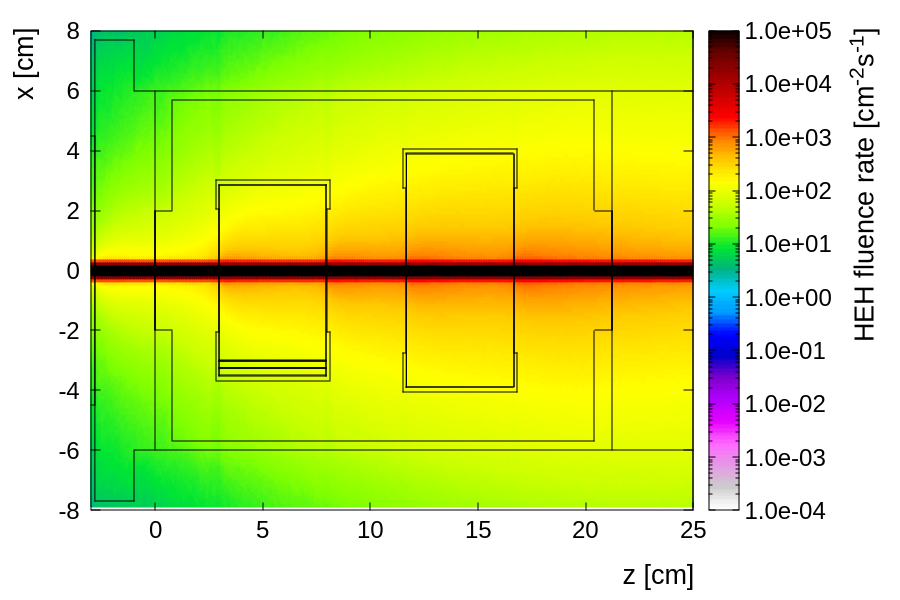}  
            \caption{\label{subfig:maphehC}}
        \end{subfigure}
    \caption{~\label{fig:mapheh} Maps of fluence rates of hadrons with energy greater than 20 MeV for U-ion (a) and C-ion (b) beams.}
\end{figure}

\begin{figure}[htbp] 
        \centering
         \begin{subfigure}{0.48\textwidth}
            \centering
            \includegraphics[width=\textwidth]{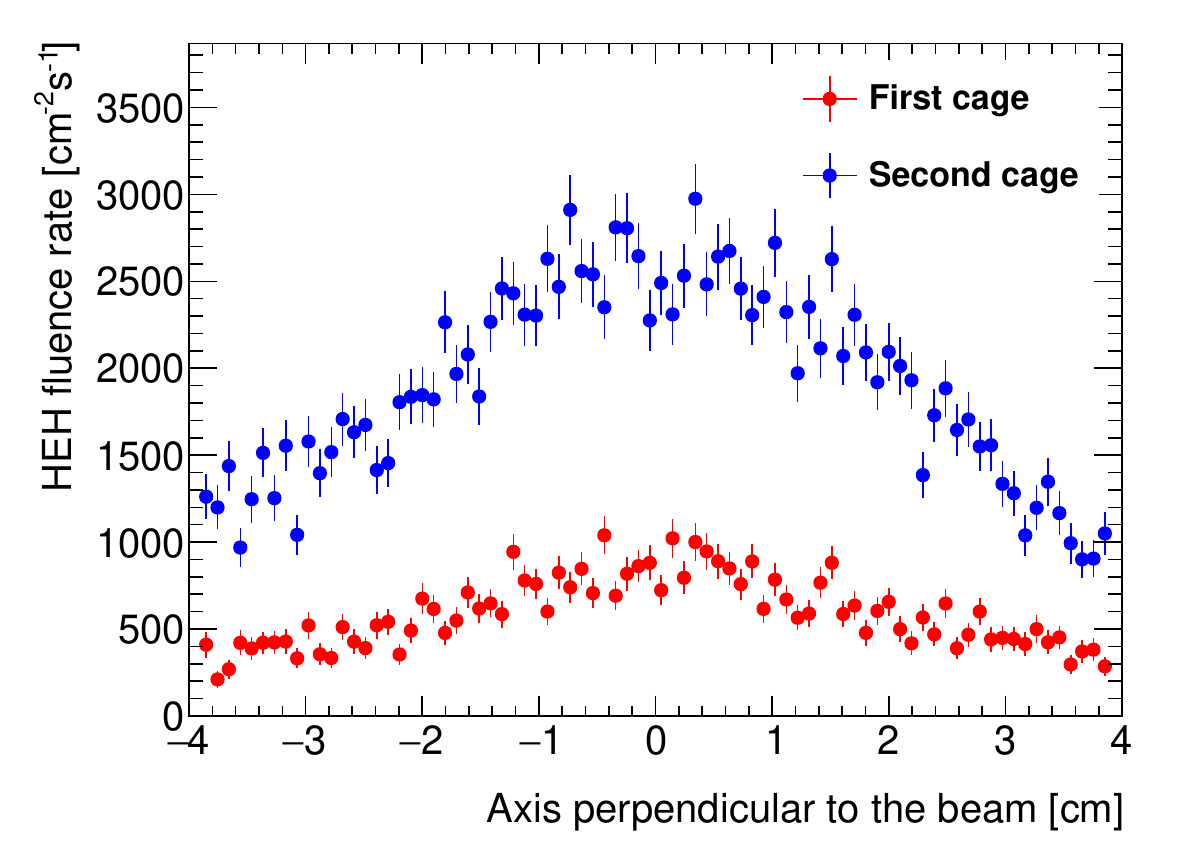}
            \caption{\label{subfig:chiphehU}}
        \end{subfigure} 
        \quad
        \begin{subfigure}{0.48\textwidth}
            \centering
            \includegraphics[width=\textwidth]{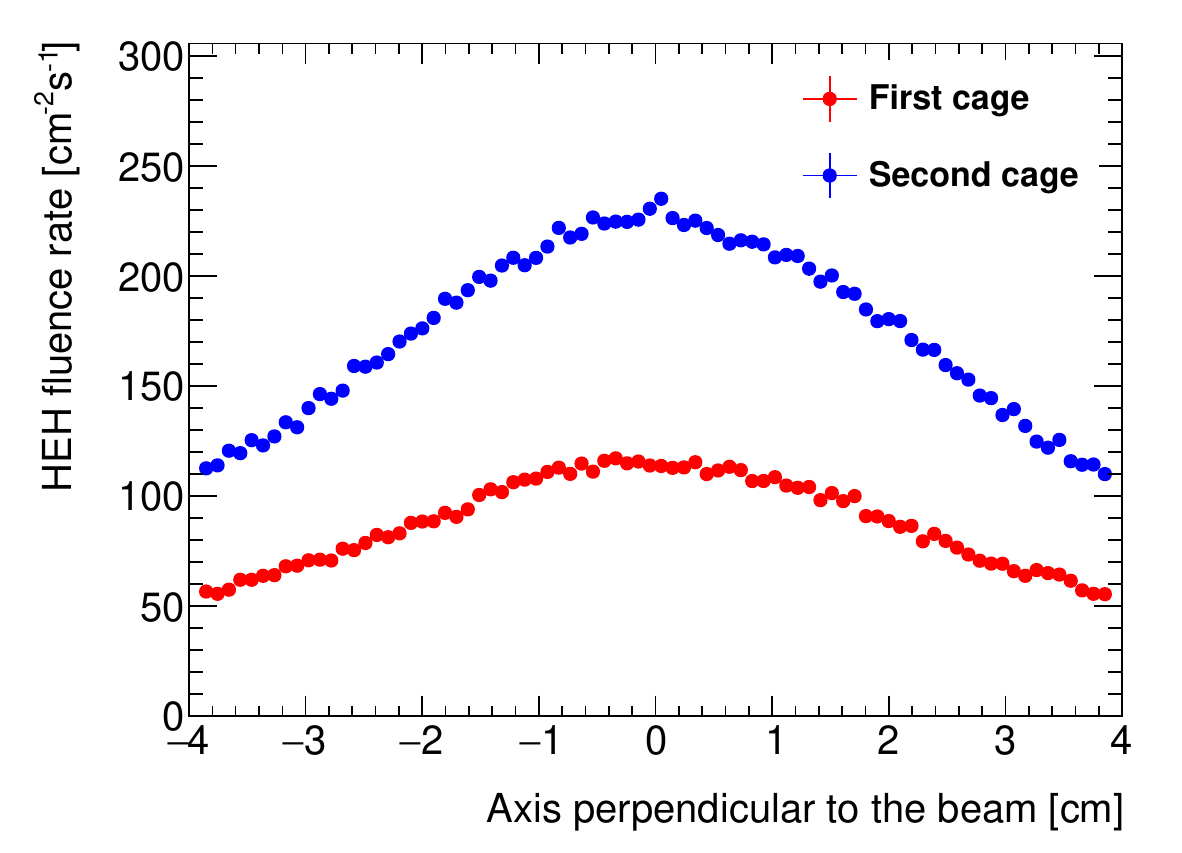}  
            \caption{\label{subfig:chiphehC}}
        \end{subfigure}
    \caption{~\label{fig:chipheh} Fluence rates of hadrons with energy greater than 20 MeV in chip planes as a function of the coordinate of axis perpendicular to the beam, for U-ion (a) and C-ion (b) beams.}
\end{figure}

The fluence rates of hadrons with energy greater than 20 MeV are shown in Figure~\ref{fig:mapheh} for the maps of BM, and in Figure~\ref{fig:chipheh} for the chip planes.
The leading contributions in the chip planes are from neutrons, which are approximately $55\%$ to $60\%$.
The subleading contributions are from protons, which are approximately $35\%$ to $40\%$.

The values for the chip planes are summarized in Table~\ref{tab:heh}.
The maximum fluence rate is 3.0 kHz/cm$^{2}$, corresponding to the chip plane in the second field cage for the U-ion beam.

\begin{table}[hbtp]
    \centering
    \caption{\label{tab:heh} The fluence rates of hadrons with energy greater than 20 MeV.
    The numbers correspond to the average values while the numbers in brackets refer to the peak values.}
    \begin{tabular}{lcc}
    \hline\hline
     & First cage [kHz/cm$^{2}$] & Second cage [kHz/cm$^{2}$] \\
    \hline
    U-ion   & $0.591 \pm 0.009$ ($1.039 \pm 0.108$) & $1.919 \pm 0.018$ ($2.975 \pm 0.201$)  \\
    C-ion   & $0.886 \pm 0.002$ ($1.171 \pm 0.019$) $\times 10^{-1}$ & $1.763 \pm 0.003$ ($2.352 \pm 0.003$) $\times 10^{-1}$ \\
    \hline\hline
    \end{tabular}
\end{table}

\subsection{Thermal neutron flux}
\label{subsec:thermaln}

\begin{figure}[htbp]
    \centering
    \includegraphics[width=0.6\textwidth]{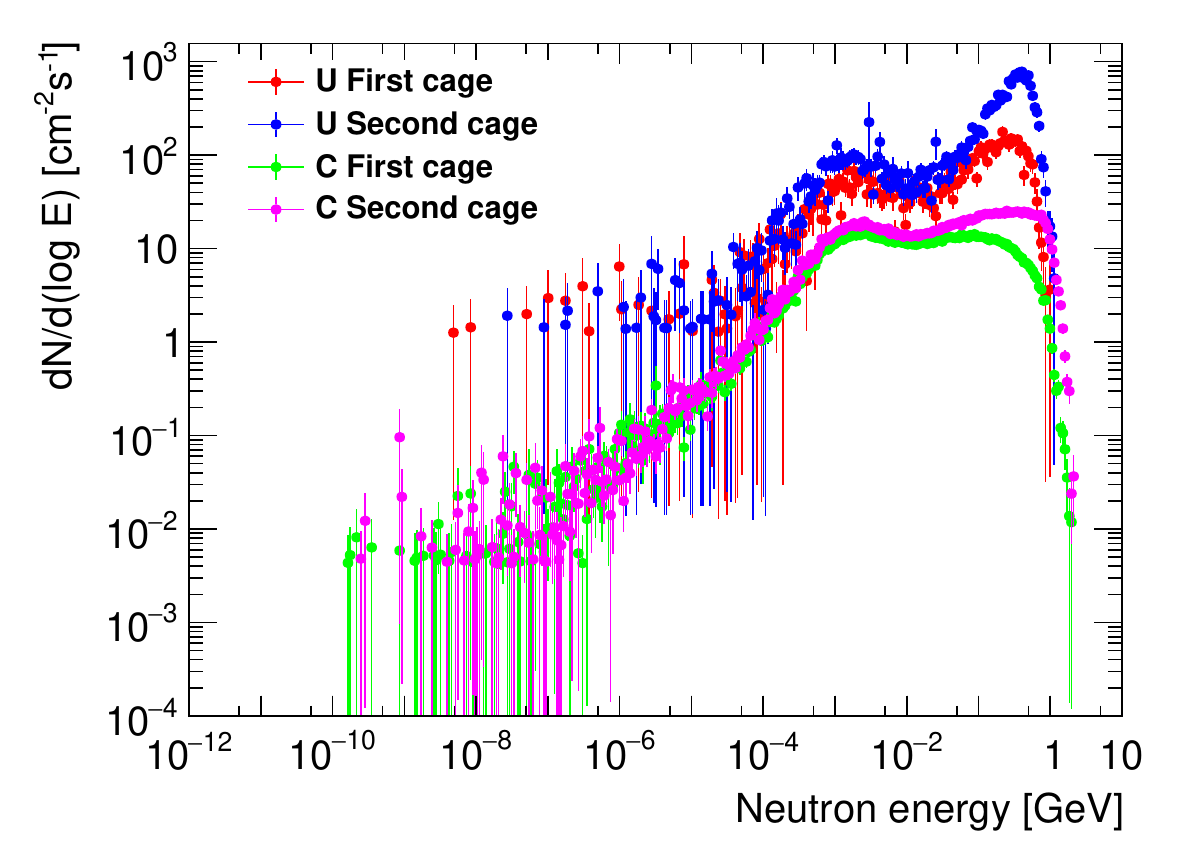}
    \caption{\label{fig:neutron_lethargy_spectra} Neutron lethargy spectra in the chip planes.}
\end{figure}

The energy spectra of neutrons averaged over the chip planes are shown in Figure~\ref{fig:neutron_lethargy_spectra} as a lethargy plot 
(i.e., showing the differential neutron fluence rate per unit logarithm of the energy, with both axes in logarithmic scale).
The thermal neutron fluence rates (with $\mathrm{E < 0.5}$ eV) for the chip planes are summarized in Table~\ref{tab:thermalneu}.

For the C-ion beam, the thermal neutron fluence rates are approximately five orders of magnitude lower than the respective high-energy hadron fluence rates.
For the U-ion beam, since no thermal neutrons are found in the simulated events, the upper limits at 95\% confidence level are given,
assuming a track length equal to the thickness of chip plane.
The limits are more than three orders of magnitude lower than the corresponding high-energy hadron fluence rates for U-ion beam.

\begin{table}[hbtp]
    \centering
    \caption{\label{tab:thermalneu} The fluence rates of thermal neutrons ($\mathrm{E < 0.5}$ eV).
    The numbers correspond to the average values over the chip planes.
    In case of zero value, upper limit at 95\% confidence level is given.}
    \begin{tabular}{lcc}
    \hline\hline
     & First cage [cm$^{-2}$s$^{-1}$] & Second cage [cm$^{-2}$s$^{-1}$] \\
    \hline
    U-ion   & $<0.26$ & $<0.26$  \\
    C-ion   & 0.0016 $\pm$ 0.0008 & 0.0012 $\pm$ 0.0009  \\
    \hline\hline
    \end{tabular}
\end{table}

\subsection{Nuclear fragment flux}
\label{subsec:frag}

\begin{figure}[htbp] 
        \centering
         \begin{subfigure}{0.48\textwidth}
            \centering
            \includegraphics[width=\textwidth]{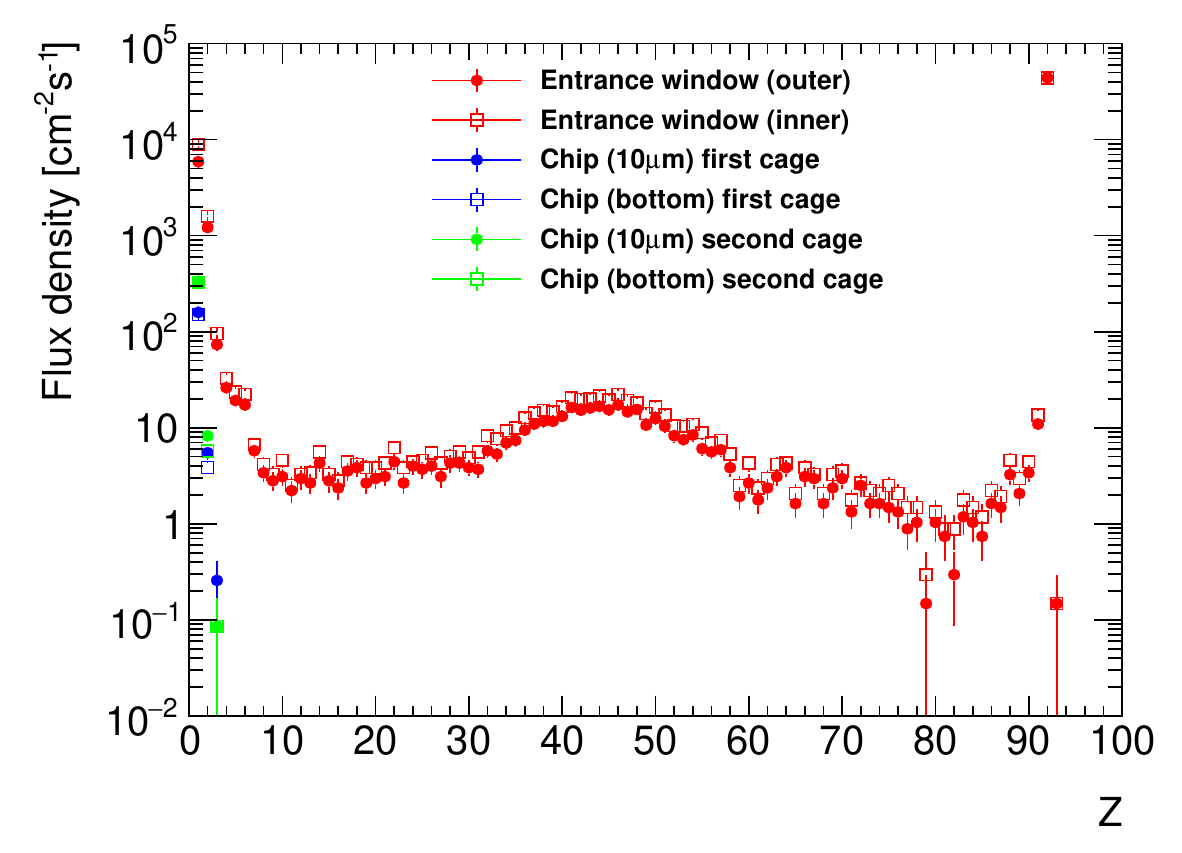}
            \caption{}
            \label{subfig:U_Z}
        \end{subfigure} 
        \quad
        \begin{subfigure}{0.48\textwidth}
            \centering
            \includegraphics[width=\textwidth]{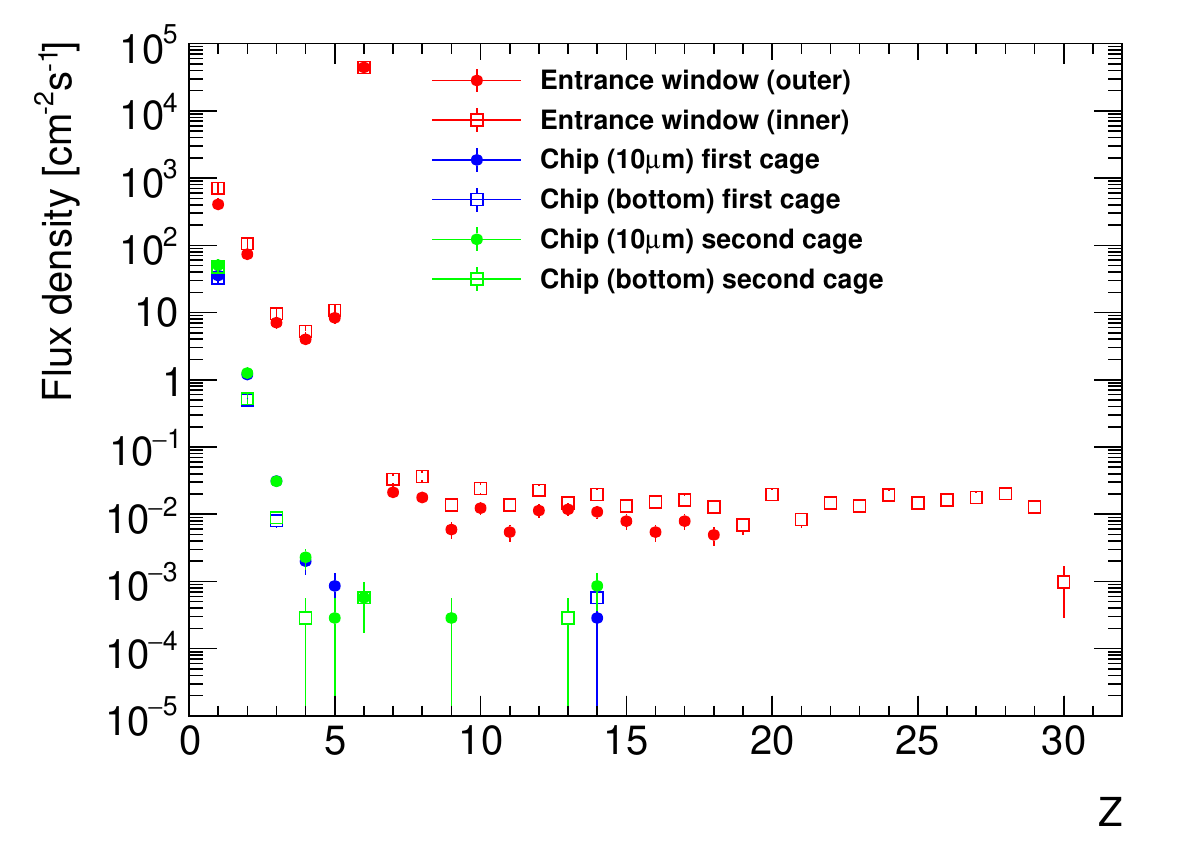} 
            \caption{}
            \label{subfig:C_Z}
        \end{subfigure}\\
        \begin{subfigure}{0.48\textwidth}
            \centering
            \includegraphics[width=\textwidth]{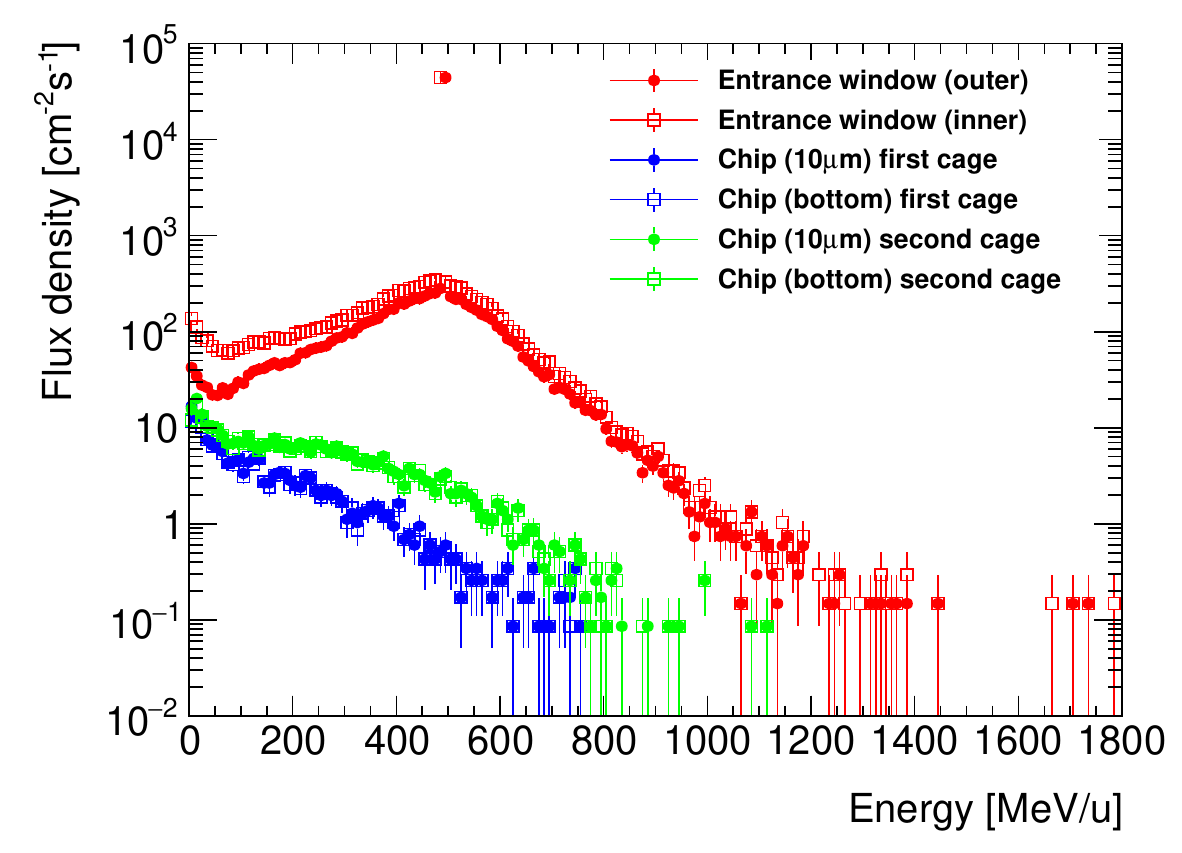}
            \caption{}
            \label{subfig:U_E}
        \end{subfigure}
        \quad
        \begin{subfigure}{0.48\textwidth}
            \centering
            \includegraphics[width=\textwidth]{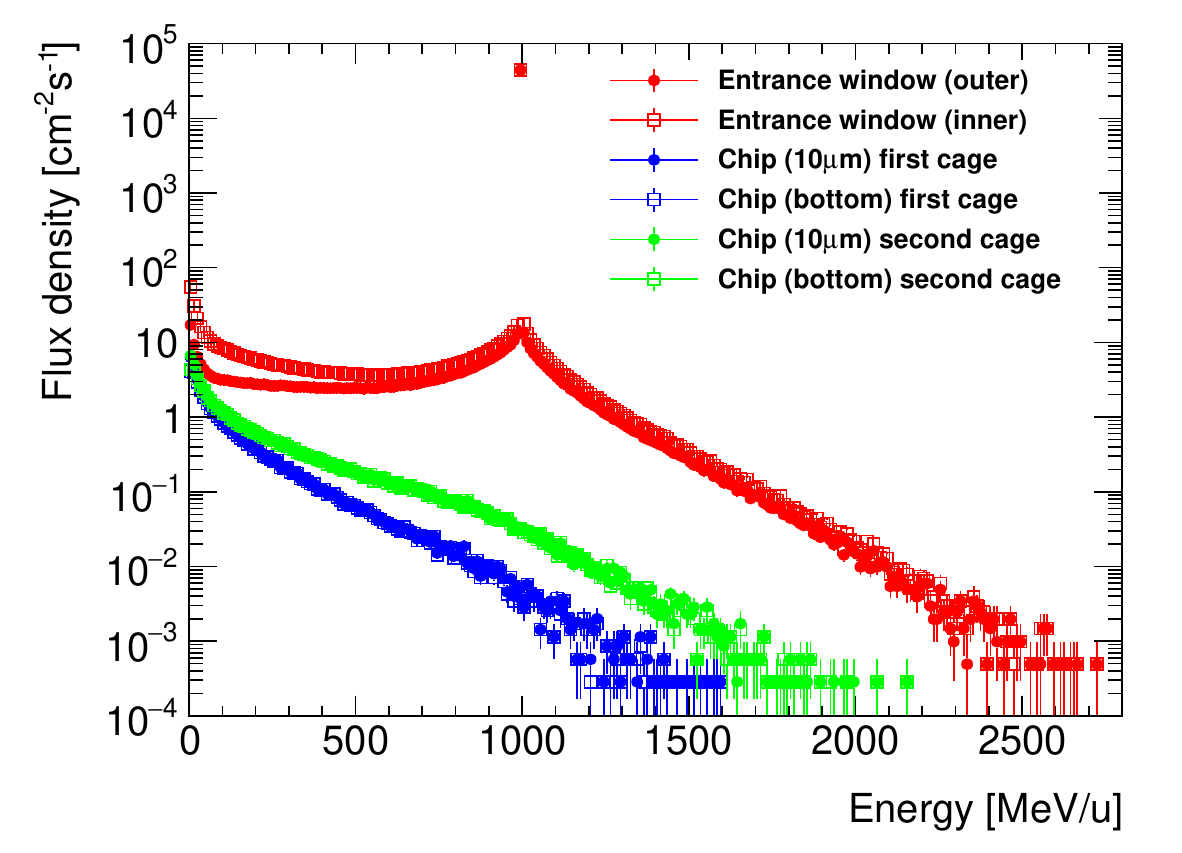}
            \caption{}
            \label{subfig:C_E}
        \end{subfigure}
        \\
        \begin{subfigure}{0.48\textwidth}
            \centering
            \includegraphics[width=\textwidth]{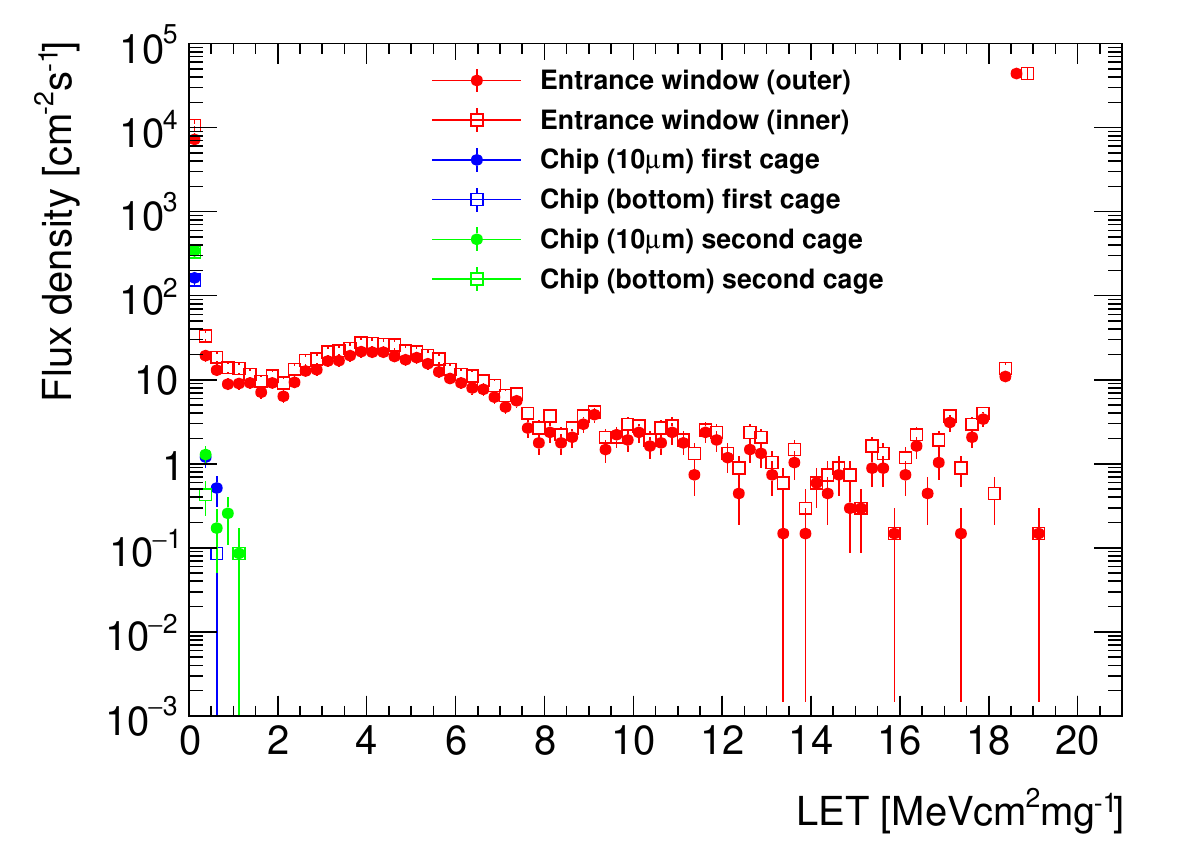}
            \caption{}
            \label{subfig:U_LET}
        \end{subfigure}
        \quad
        \begin{subfigure}{0.48\textwidth}
            \centering
            \includegraphics[width=\textwidth]{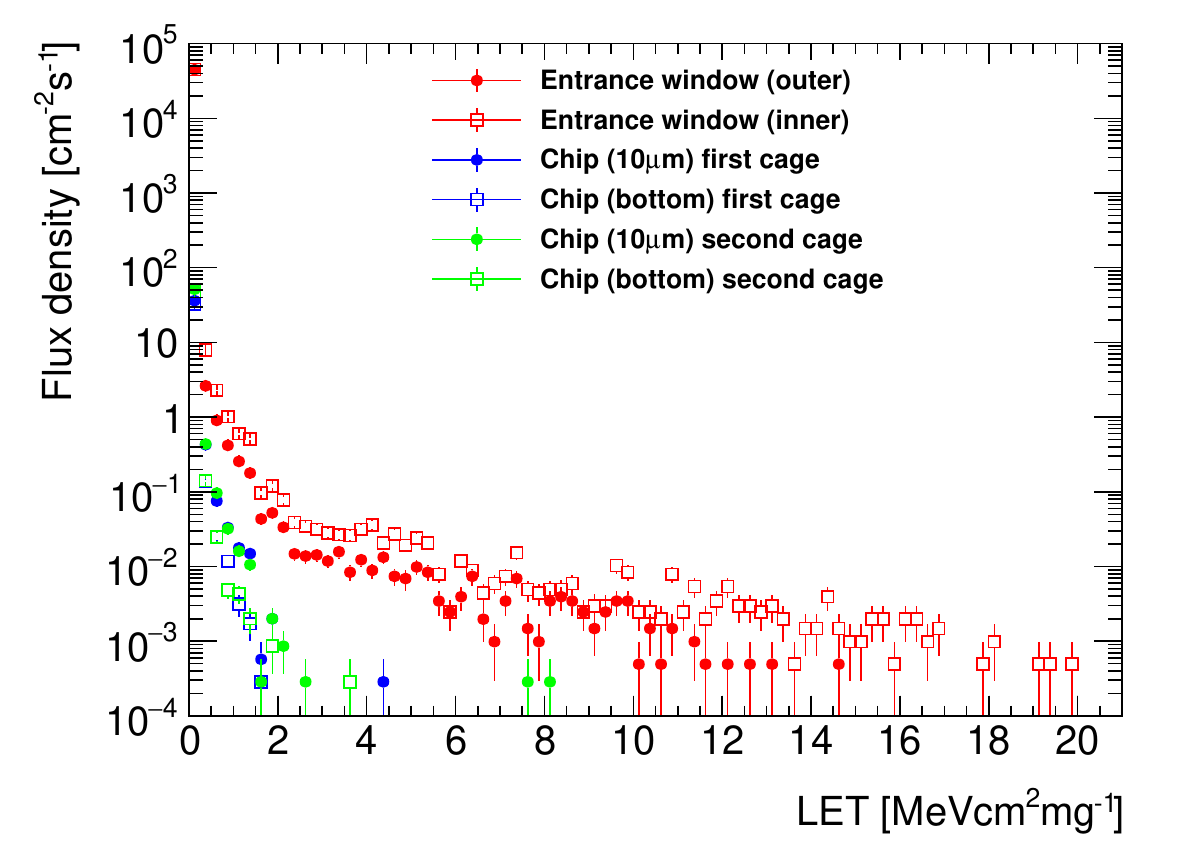}
            \caption{}
            \label{subfig:C_LET}
        \end{subfigure}
    \caption{\label{fig:frag} Flux densities of nuclear fragments as a function of atomic number Z for U-ion (a) and C-ion (b) beams, 
                              as a function of energy for U-ion (c) and C-ion (d) beams, 
                              and as a function of LET for U-ion (e) and C-ion (f) beams.}
\end{figure}

The nuclear fragments can carry large kinetic energies per nucleon close to the incident particles,
or are of small kinetic energies (i.e. target-like fragments).
Six surfaces are studied to evaluate the flux densities of the nuclear fragments: 10 $\mu$m below the top of the two chip planes, the bottom of the two chip planes, the outer and inner sides of the entrance window of the second field cage.
The surface at 10 $\mu$m below the top is about the sensitive area of the chip.
Figure~\ref{fig:frag} shows the flux densities of the nuclear fragments as a function of atomic number Z, as a function of energy, and as a function of LET for U-ion and C-ion beams.
Table~\ref{tab:frag} summarizes the flux densities of the nuclear fragments with Z $\geq$ 2, and of those with LET $>$ 0.25 MeVcm$^2$mg$^{-1}$ (i.e. excluding the first bin in Figure~\ref{subfig:U_LET} and~~\ref{subfig:C_LET}), for the two kinds of surfaces in the chip planes.

For all the plots, the peaks corresponding to the U ions or C ions of the beams are visible for the surfaces of the entrance window.
For fragments with Z $\geq$ 2 (LET $>$ 0.25 MeVcm$^2$mg$^{-1}$), the flux densities at 10 $\mu$m below the top of the chip planes are about 1.5 to 2.5 (3.0 to 3.5) times higher than
the corresponding values at the bottom of chip planes.

\begin{table}[hbtp]
    \centering
    \caption{\label{tab:frag} The flux densities of the nuclear fragments with Z $\geq$ 2, and of those with LET $>$ 0.25 MeVcm$^2$mg$^{-1}$.
    Results are presented for the surfaces at 10 $\mu$m below the top of the two chip planes, and at the bottom of them.}
    \begin{tabular}{lcccc}
    \hline\hline
     & \multicolumn{2}{c}{First cage [cm$^{-2}$s$^{-1}$]} & \multicolumn{2}{c}{Second cage [cm$^{-2}$s$^{-1}$]} \\
     &  10 $\mu$m  & bottom & 10 $\mu$m  & bottom \\ 
    \hline
    \multicolumn{5}{c}{Z $\geq$ 2} \\
    U-ion   & $5.7 \pm 0.7$ & $3.9 \pm 0.6$ & $8.3 \pm 0.9$ & $5.8 \pm 0.7$ \\
    C-ion   & $1.23 \pm 0.02$ & $0.50 \pm 0.01$ & $1.29 \pm 0.02$ & $0.52 \pm 0.01$ \\
    \hline
    \multicolumn{5}{c}{LET $>$ 0.25 MeVcm$^2$mg$^{-1}$} \\
    U-ion   & $1.7 \pm 0.4$ & $0.5 \pm 0.2$ & $1.8 \pm 0.4$ & $0.5 \pm 0.2$ \\
    C-ion   & $0.57 \pm 0.01$ & $0.181 \pm 0.007$ & $0.60 \pm 0.01$ & $0.178 \pm 0.008$ \\
    \hline\hline
    \end{tabular}
\end{table}

\subsection{Alternative simulation setups}
\label{subsec:alter}

In Section~\ref{subsubsec:shield} to~\ref{subsubsec:detposition}, simulations with alternative geometries, detector materials and beam conditions are described.
The results and discussions are collectively presented in Section~\ref{subsubsec:result}.

\subsubsection{Shielding}
\label{subsubsec:shield}

\begin{figure}[htbp]
    \centering
    \includegraphics[width=0.46\textwidth]{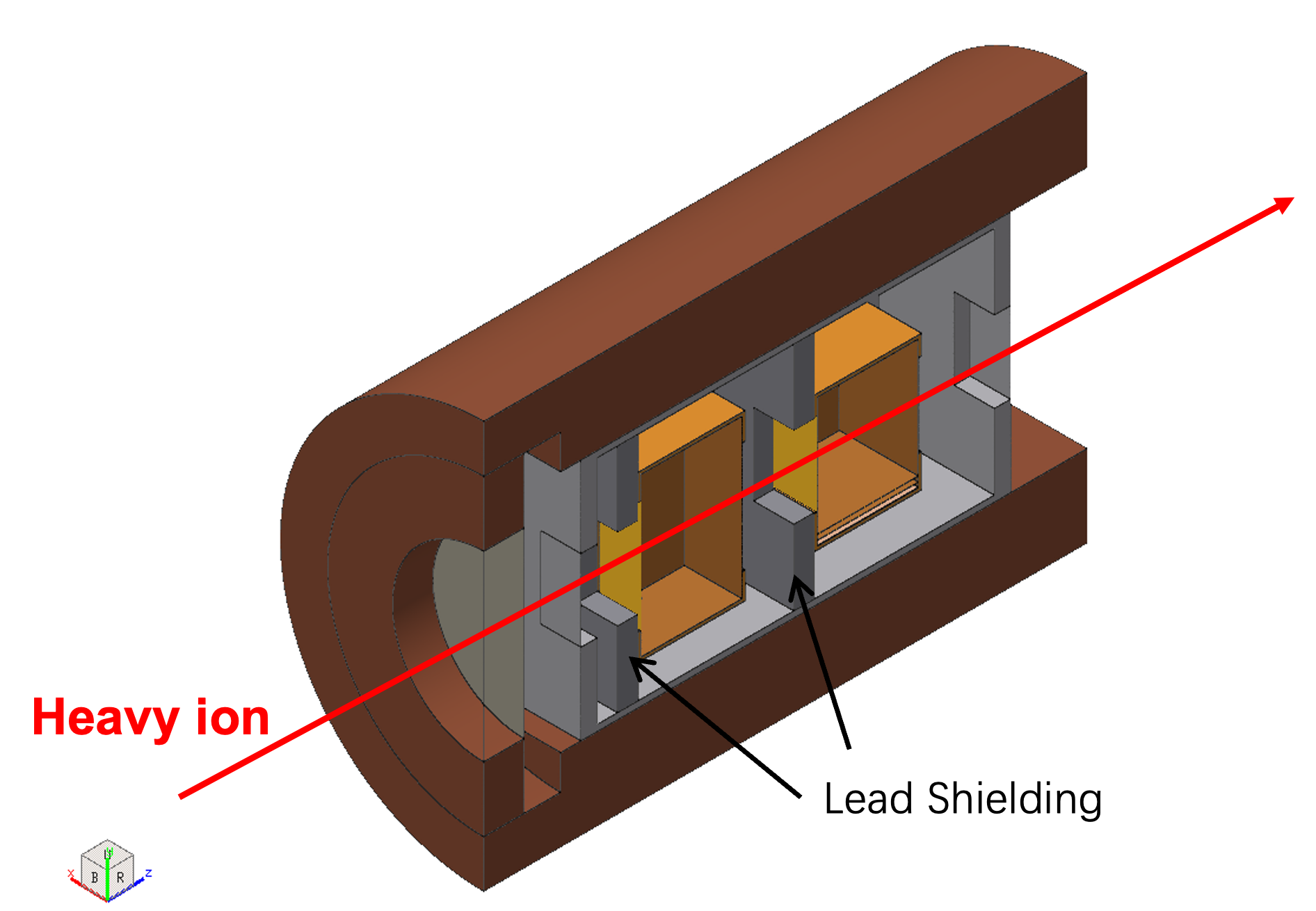}
    \caption{The shielding scheme of BM.}\label{fig:Shielding_BM}
\end{figure}

The shielding is investigated to reduce the radiation levels on the chip planes.
As there are empty areas before the field cages inside the gas chamber, lead plates of approximately 1 cm thick could be placed there.
Figure~\ref{fig:Shielding_BM} shows the shielding scheme of BM.
For each lead plate, a 40 ($x$) $\times$ 40 ($y$) mm$^2$ hole is left out for the beam. 

\subsubsection{Field cage with less material} 
\label{subsubsec:cagev1}

To reduce the material budget of BM in the path of the beam, the field cages with thinner entrance/exit windows have been developed and are being tested.
The 25.4 $\mu$m thick Kapton films are replaced by 2 $\mu$m thick Mylar films.
And the drift electrodes of 35 $\mu$m thick copper cladded with 0.05 $\mu$m thick gold are replaced by 100 nm thick aluminum.

\subsubsection{Number of GEM layers}
\label{subsubsec:gem}

According to the performance of the Topmetal-CEE chips~\cite{LIU2023167786} and readout electronics~\cite{Liu2025},
two to zero GEM layers are to be used for the ion species from C to U.
The pitch of the pixels in the chip is 100 $\mu$m, which is finer than the 150 $\mu$m pitch of the holes in the GEM.
Performance-wise, without considering the effect of radiation on the electronics of BM, 
smaller number of GEM layers are preferred, to achieve better position and time resolutions.

\subsubsection{Beam widths}
\label{subsubsec:beamwidth}

To study the dependence of radiation level on the beam width, the beams with FWHMs of 1 mm, 4 mm and 6 mm are simulated, in addition to the benchmark value of 2.35 mm.

\subsubsection{Detector position}
\label{subsubsec:detposition}

There is a possibility that the distance between the beam exit window and the BM is increased from 2.8 cm to 20 cm. 
Its effect is studied.

\subsubsection{Results and discussions}
\label{subsubsec:result}

Table~\ref{tab:altersetup} summarizes the TIDs, NIELs in 1 MeV neutron equivalent fluences, and fluence rates of hadrons with energy greater than 20 MeV, for various simulation setups in case of U-ion beam.

\begin{table}[hbtp]
    \centering
    \caption{\label{tab:altersetup} Results of various setups for U-ion beam.}
  \resizebox{\textwidth}{!}{
    \begin{tabular}{lcccccc}
    \hline\hline
             & \multicolumn{2}{c}{TID} & \multicolumn{2}{c}{1-MeV-neq} &  \multicolumn{2}{c}{>20 MeV had.} \\
             & \multicolumn{2}{c}{[kGy]} & \multicolumn{2}{c}{[$\times 10^{11}$ cm$^{-2}$]} &  \multicolumn{2}{c}{[kHz/cm$^2$]} \\
    \hline
             & First cage & Second cage & First cage & Second cage & First cage & Second cage \\
    \hline
    Nominal setup & 5.699 $\pm$ 0.002 & 7.409 $\pm$ 0.002 & 0.959 $\pm$ 0.002 & 1.541 $\pm$ 0.004 & 0.592 $\pm$ 0.009 & 1.919 $\pm$ 0.018 \\
    \hline
    With shielding & 5.732 $\pm$ 0.002 & 7.194 $\pm$ 0.002 & 0.981 $\pm$ 0.003 & 1.538 $\pm$ 0.004 & 0.565 $\pm$ 0.009 & 1.803 $\pm$ 0.018 \\
    \hline
    Cage with less material & 1.655 $\pm$ 0.001 & 1.560 $\pm$ 0.001 & 0.275 $\pm$ 0.001 & 0.299 $\pm$ 0.001 & 0.342 $\pm$ 0.004 & 0.686 $\pm$ 0.007 \\
    \hline
    0 GEM layer & 13.731 $\pm$ 0.003 & 16.272 $\pm$ 0.003 & 1.540 $\pm$ 0.003 & 2.324 $\pm$ 0.004 & 0.615 $\pm$ 0.009 & 1.962 $\pm$ 0.018 \\
    1 GEM layer & 8.098 $\pm$ 0.002 & 10.138 $\pm$ 0.003 & 1.183 $\pm$ 0.003 & 1.861 $\pm$ 0.004 & 0.591 $\pm$ 0.009 & 1.946 $\pm$ 0.018 \\
    \hline
    Beam FWHM = 1 mm & 5.696 $\pm$ 0.002 & 7.409 $\pm$ 0.002 & 0.944 $\pm$ 0.002 & 1.520 $\pm$ 0.004 & 0.536 $\pm$ 0.009 & 1.741 $\pm$ 0.017 \\
    Beam FWHM = 4 mm & 5.704 $\pm$ 0.002 & 7.407 $\pm$ 0.002 & 0.959 $\pm$ 0.003 & 1.544 $\pm$ 0.004 & 0.594 $\pm$ 0.009 & 1.935 $\pm$ 0.018 \\
    Beam FWHM = 6 mm & 5.712 $\pm$ 0.002 & 7.406 $\pm$ 0.002 & 0.962 $\pm$ 0.003 & 1.554 $\pm$ 0.004 & 0.612 $\pm$ 0.009 & 1.973 $\pm$ 0.018 \\
    \hline
    BM position & 5.434 $\pm$ 0.002 & 7.388 $\pm$ 0.002 & 0.985 $\pm$ 0.003 & 1.616 $\pm$ 0.004 & 1.194 $\pm$ 0.015 & 2.492 $\pm$ 0.022 \\
    \hline\hline
    \end{tabular}
  }
\end{table}

The shielding layers have no visible impact on TIDs, NIELs, or high energy hadron fluence rates.
The reason is that the electrons contributing to the TIDs and NIELs are mainly produced in the gas inside the field cages or in their entrance windows,
while the 1 cm thick lead can not shield the protons and neutrons with energies of several hundreds MeV.

The cages with thinner entrance/exit windows result in smaller radiation levels. 
Both TIDs and NIELs are reduced to approximately $30\%$ ($20\%$) of the nominal values for the chip planes in the first (second) field cages,
while the corresponding numbers are approximately $60\%$ ($35\%$) for high-energy hadron fluence rates.

\begin{figure}[htbp]
    \centering
    \includegraphics[width=0.48\textwidth]{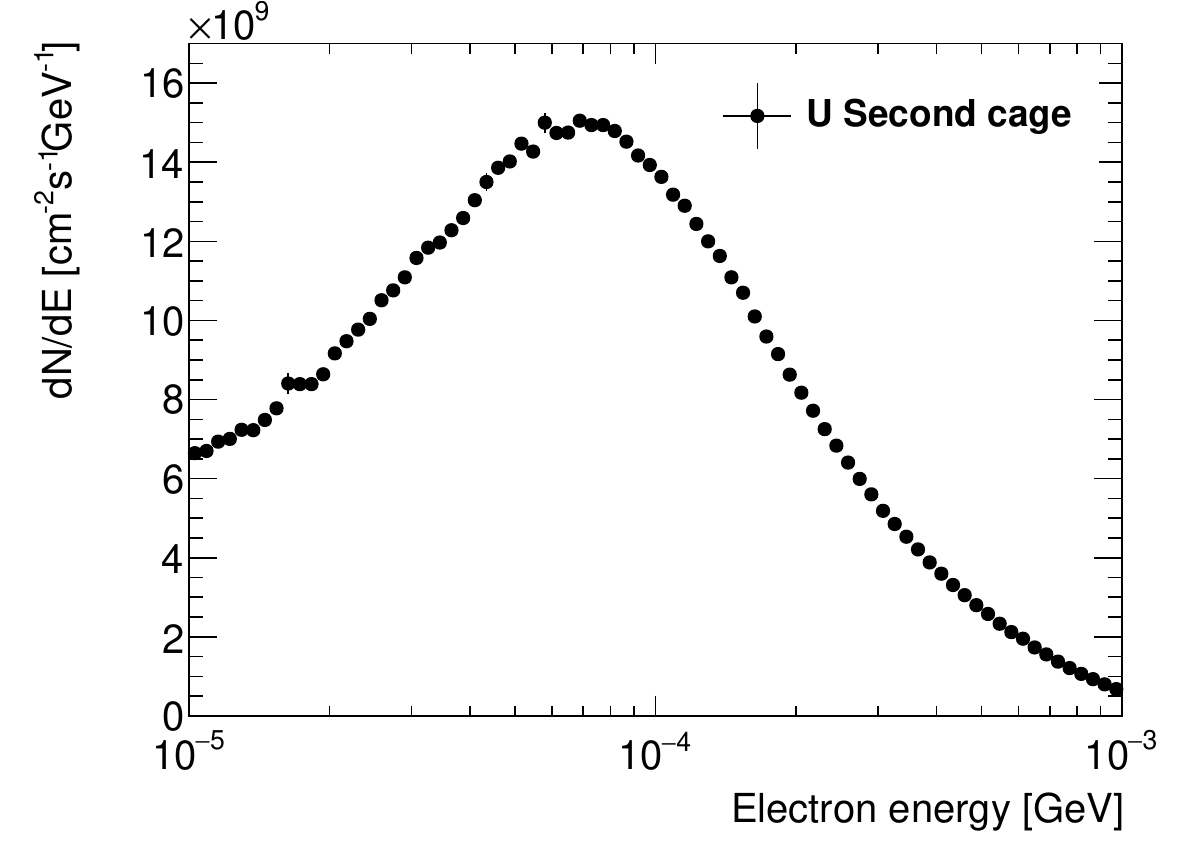}
    \caption{\label{fig:elecspectra} The energy spectrum of electrons reaching the top surface of the chip plane in the second field cage with zero GEM layers for U-ion beam.}
\end{figure}

Adding two GEM layers reduces the TIDs by approximately $55\%$, and reduces the NIELs by approximately $35\%$. 
The energy spectrum of electrons reaching the top surface of the chip plane is shown in Figure~\ref{fig:elecspectra} for the case of zero GEM layers.
The peak value is at about 70 keV.
The GEM layers have no significant impact on the high-energy hadron fluence rates, given the high penetration powers of high-energy hadrons.

The dependences on the beam width are weak for all three radiation quantities.
As the FWHM of beam increases from 1 mm to 6 mm, the TIDs are almost unchanged, while the NIELs and high-energy hadron fluence rates increase by approximately $2\%$ and $13\%$, respectively.

Increased distance between the beam exit window and BM has no significant impact on the TIDs and NIELs.
It results in increased high-energy hadron fluence rates, due to the increased materials and high penetration power of high-energy hadrons.
The high-energy hadron fluence rates for the chip planes in the first and second cages increase by $100\%$ and $30\%$, respectively.

\section{Conclusion}
\label{sec:conc}

In this paper, the radiation environment for the BM of CEE experiment is presented, in particular for the chip planes.
The studied radiation quantities include TID, NIEL expressed in 1 MeV neutron equivalent fluence, high-energy hadron flux, thermal neutron flux, and nuclear fragment flux.
The maximum TID for the chip planes is approximately 10.6 kGy, 
the maximum 1 MeV neutron equivalent fluence is approximately 2.2 $\times 10^{11}$ cm$^{-2}$,
and the maximum fluence rate of hadrons with energy greater than 20 MeV is approximately 3.0 kHz/cm$^{2}$.
Alternative simulation setups are also investigated, including possible shielding, thinner entrance/exit windows of field cages, 
different number of GEM layers, various beam widths, longer distance between the beam exit window and BM.

This study provides the required numbers for the assessment of expected performance of Topmetal-CEE sensor chips in the radiation environment of CEE experiment.
Irradiation tests of Topmetal-CEE chips are currently being performed.
In addition, the study of alternative simulation setups provides references for the operation strategies of BM.

\acknowledgments

This work was supported by the National Key Research and Development Program of China (No. 2024YFA1610700, 2024YFE0110101), 
the National Natural Science Foundation of China (Nos. 11927901, 12105110, 12275105), 
Guizhou Science and Technology Foundation-ZK [2023] General (No. 248).


\bibliographystyle{JHEP}
\bibliography{PaperBMradiation.bib}






\end{document}